\documentclass[a4paper,11pt]{article}

\usepackage{a4wide}

\usepackage{times}

\usepackage[latin1]{inputenc}
\usepackage{amsfonts}
\usepackage{amsmath}
\usepackage{amssymb}
\usepackage{amsthm}
\usepackage[american]{babel}
\usepackage{fontenc}
\usepackage{graphicx}
\usepackage[normal]{subfigure}

\providecommand{\I}{\mathcal{I}}

\providecommand{\setZ}{\mathbb{Z}}
\providecommand{\setN}{\mathbb{Z}}

\newtheoremstyle{theorem}{1em}{1em}{\slshape}{0pt}{\bfseries}{.}{ }{}
\theoremstyle{theorem}
\newtheorem{theorem}{Theorem}
\newtheorem{claim}[theorem]{Claim}
\newtheorem{definition}[theorem]{Definition}
\newtheorem{lemma}[theorem]{Lemma}
\newtheorem{property}[theorem]{Property}

\author{Jaroslaw Byrka \and Andreas Karrenbauer\thanks{supported by the Deutsche Forschungsgemeinschaft (DFG)  within  Priority Programme 1307  ``Algorithm Engineering".}
 \and Laura Sanit\`a\thanks{supported by Swiss National Science Foundation within the project  ``Robust Network Design".} \\ \\
Institute of Mathematics \\ 
					EPFL, Lausanne, Switzerland \\
					{\{Jaroslaw.Byrka,Andreas.Karrenbauer,Laura.Sanita\}@epfl.ch}
}
\date{}

\title{The interval constrained 3-coloring problem}

\begin{document}
\maketitle

\begin{abstract}
In this paper, we settle the open complexity status of interval constrained coloring with a fixed number of colors. We prove that the problem is already NP-complete if the number of different colors is 3. Previously, it has only been known that it is NP-complete, if the number of colors is part of the input and that the problem is solvable in polynomial time, if the number of colors is at most 2. We also show that it is hard to satisfy almost all of the constraints for a feasible instance. This implies APX-hardness of maximizing the number of simultaneously satisfiable intervals.
\end{abstract}

\section{Introduction}

In the interval constrained 3-coloring problem, we are given a set $\I$ of intervals defined on $[n] := \{1,\ldots,n\}$ and a \emph{requirement} function $r : \I \to \setZ^3_{\ge 0}$, which maps each interval to a triple of non-negative integers. The objective is to determine a coloring $\chi : [n] \to \{1,2,3\}$ such that each interval gets the proper colors as specified by the requirements, i.e.~$\sum_{i \in I} e_{\chi(i)} = r(I)$ where $e_1,e_2,e_3$ are the three unit vectors of $\setZ^3$.

This problem is motivated by an application in biochemistry to investigate the tertiary structure of proteins as shown in the following illustration.
\begin{figure}[hb]
 \centering
 \includegraphics[height=30mm,page=1]{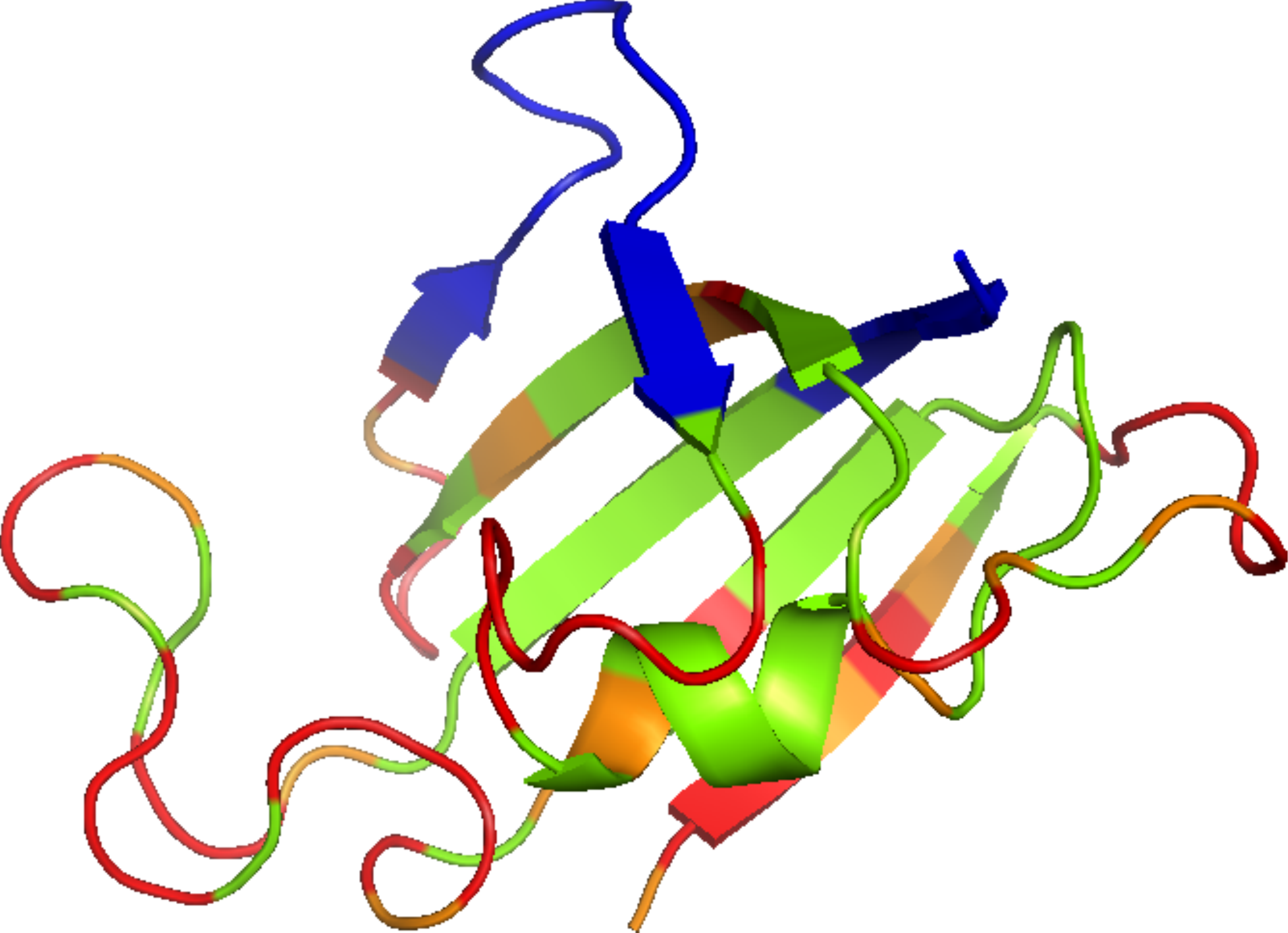}
 \caption{Coloring of the residues of a protein chain according to their exchange rates.}
 \label{fig:bio}
\end{figure}
More precisely, in Hydrogen-Deuterium-Exchange (HDX) experiments proteins are put into a solvent of heavy water ($D_2O$) for a certain time after which the amount of residual hydrogen atoms, that have exchanged with deuterium atoms, is measured~\cite{LEHMP02}. Doing this experiment for several timesteps, one can determine the exchange rate of the residues. These exchange rates indicate the solvent accessibility of the residues and hence they provide information about the spatial structure of the protein. Mass spectroscopy is one of the methods for measuring these exchange rates. To this end, the proteins are digested, i.e.~cut into parts which can be considered as intervals of the protein chain, and the mass uptake of each interval is measured. But thereby only bulk information about each interval can be obtained. Since there is not only one protein in the solvent but millions and they are not always cut in the same manner, we have this bulk information on overlapping fragments. That is, we are given the number of slow, medium, and fast exchanging residues for each of these intervals and our goal is to find a feasible assignment of these three exchange rates to residues such that for each interval the numbers match with the bulk information. 

Though the interval constrained 3-coloring problem is motivated by a particular application, its mathematical abstraction appears quite simple and ostensibly more general. In terms of integer linear programming, the problem can be equivalently formulated as follows. Given a matrix $A \in \{0,1\}^{m \times n}$ with the \emph{row-wise consecutive-ones property} and three vectors $b_{1,2,3} \in \setZ_{\ge 0}^m$, the constraints
\begin{equation}
\label{eq}
\begin{pmatrix} 
A & 0 & 0 \\
0 & A & 0 \\
0 & 0 & A \\
I & I & I
\end{pmatrix}
\cdot
\begin{pmatrix}
x_1 \\ x_2 \\ x_3  
\end{pmatrix}
=
\begin{pmatrix}
b_1 \\ b_2 \\ b_3 \\ 1
\end{pmatrix}
\end{equation}
have a binary solution, i.e.~$x_{1,2,3} \in \{0,1\}^n$, if and only if the corresponding interval constrained 3-coloring problem has a feasible solution. We may assume w.l.o.g.~that the requirements are consistent with the interval lengths, i.e.~$A\cdot 1 = b_1 + b_2 + b_3$, since otherwise we can easily reject the instance as infeasible. Hence, we could treat $x_3$ as slack variables and reformulate the constraints as
\begin{equation}\label{eq:packing}
 A x_1 = b_1, \qquad A x_2 = b_2, \qquad x_1 + x_2 \leq 1.
\end{equation}
It is known that if the matrix $A$ has the \emph{column-wise} consecutive ones property (instead of \emph{row-wise}), then there is a reduction from the two-commodity integral flow problem, which has been proven to be NP-complete in~\cite{EIS76}. However, the NP-completeness w.r.t.~row-wise consecutive ones matrices has been an open problem in a series of papers as outlined in the following subsection. 

\subsection{Related Work}

The problem of assigning exchange rates to single residues has first been considered in~\cite{SAC08}. In that paper, the authors presented a branch-and-bound framework for solving the corresponding coloring problem with $k$ color classes. They showed that there is a combinatorial polynomial time algorithm for the case of $k=2$. Moreover, they asked the question about the complexity for $k > 2$. In~\cite{SWAT08}, the problem has been called \emph{interval constrained coloring}. It has been shown that the problem is NP-hard if the parameter $k$ is part of the input. Moreover, approximation algorithms have been presented that allow violations of the requirements. That is, a quasi-polynomial time algorithm that computes a solution in which all constraints are $(1+\varepsilon)$-satisfied and a polynomial time rounding scheme, which satisfies every requirement within $\pm 1$, based on a technique introduced in~\cite{GKPS06}. The latter implies that if the LP relaxation of~\eqref{eq} is feasible, then there is a coloring satisfying at least $\tfrac{5}{16}$ of the requirements. APX-hardness of finding the maximum number of simultaneously satisfiable intervals has been shown in~\cite{Can08} for $k \ge 2$ provided that intervals may be counted with multiplicities. But still, the question about the complexity of the decision problem for fixed $k \geq 3$ has been left open. In~\cite{KNU09}, several fixed parameter tractability results have been given. However, the authors state that they do not know whether the problem is tractable for fixed $k$. 

\subsection{Our contribution}

In this paper, we prove the hardness of the interval constrained $k$-coloring problem for fixed parameter $k$. In fact, we completely settle the complexity status of the problem, since 
we show that already the interval constrained 3-coloring problem is NP-hard by a reduction from 3-SAT. This hardness result holds more generally for any problem that can be formulated like \eqref{eq}.
Moreover, we even show the stronger result, that it is still difficult to satisfy almost all of the constraints for a feasible instance. More precisely, we prove that there is a constant $\epsilon > 0$ such that it is NP-hard to distinguish between instances where all constraints can be satisfied and those where only a $(1-\epsilon)$ fraction of constraints can be simultaneously satisfied. To this end, we extend our reduction using expander graphs. This gap hardness result implies APX-hardness of the problem of maximizing the number of satisfied constraints. It is important to note that our construction does neither rely on multiple copies of intervals nor on inconsistent requirements for an interval, i.e.~in our construction for every interval $(i,j)$ we have unique requirements that sum up to the length of the interval.

\section{NP-hardness}

\begin{theorem}
It is NP-hard to decide whether there exists a feasible coloring $\chi$ for an instance $(\I, r)$ of the interval constrained 3-coloring problem.
\end{theorem}

\begin{proof}
The proof is by reduction from the 3-SAT problem.

\smallskip
\noindent Suppose to be given an instance of the 3-SAT problem, defined by $q$ clauses $C_1, \dots, C_q$ and $p$ variables $x_1, \dots, x_p$. Each clause $C_i$ $(i=1, \dots, q)$ contains 3 literals, namely $y_1(i),y_2(i),y_3(i)$. Each literal
$y_h(i)$ $(i=1, \dots, q$ and $h=1,2,3)$ \emph{refers to} a variable $x_j$, that means, it is equal to either $x_j$ or $\bar x_j$ for some $j$ in $1, \dots, p$. A truth assignment for the variables $x_1, \dots, x_p$ satisfies the 3-SAT instance 
if and only if, for each clause, at least one literal takes the value $true$.

\begin{figure}[htb]
\centering
\includegraphics[width = 0.875 \textwidth,page=2]{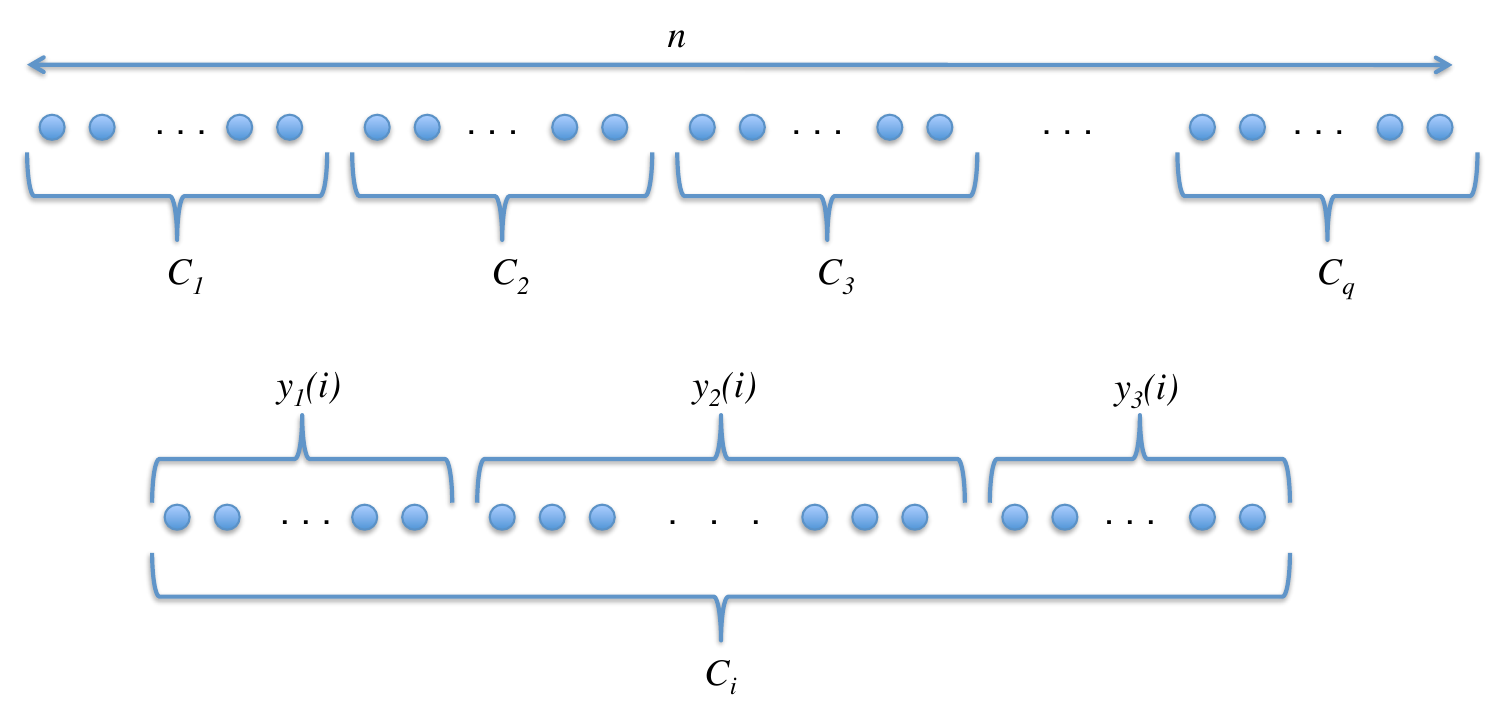} 
\caption{The sequence of nodes in an instance of the interval constrained 3-coloring problem.}
\label{fig:1}
\end{figure}

We now construct an instance of the interval constrained 3-coloring problem.
For each clause $C_i$ we introduce a sequence of consecutive nodes.
This sequence is, in its turn,  the union of three subsequences, one for each of the three literals (see Fig.~\ref{fig:1}).

In the following, for the clarity of presentation, we drop the index $i$, if it is clear from the context. We denote color 1 by RED, color 2 by BLACK and color 3 by WHITE.

\paragraph{Literal $y_1(i)$.}  The subsequence representing literal $y_1$ is composed of 8 nodes. Among them, there are three special nodes, namely $t_1,f_1$ and $a_1$, that play a key role since they encode the information about the truth value of the literal and of the variable $x_j$ it refers to. The basic idea is to achieve the following two goals: 1) given a feasible coloring, if $\chi(t_1)$ is BLACK, we want to be able to construct a truth assignment setting $x_j$ to $true$, while if $\chi(f_1)$ is BLACK, we want to be able to construct a truth assignment setting the variable $x_j$ to $false$; 2) given a feasible coloring, if $\chi(a_1)$ is RED, we want to be able to construct a truth assignment where $y_1$ is $true$.

To achieve the first goal, we will impose the following property: 

\begin{property}\label{(i)}
In any feasible coloring, exactly one among $t_1$ and $f_1$ will be BLACK.
\end{property}

\noindent To achieve the second goal, and being consistent with the first one, we must have the property that:

\begin{property}\label{(ii)}
In any feasible coloring, if $\chi(a_1) = RED$, then $\chi(t_1) = BLACK$ if $y_1 = x_j$, while $\chi(f_1) = BLACK$ if $y_1 = \bar x_j$. 
\end{property}

\noindent To guarantee properties~\eqref{(i)} and \eqref{(ii)}, we introduce a suitable set $\mathcal I(y_1)$ of 
six intervals\footnote{In principle, interval $I_5$ and the node it contains are not needed. However, this allows to have the same number of WHITE and BLACK colored nodes for the sake of exposition.}, shown in Fig.~\ref{fig:2}a. 

\begin{figure}[htb]
\centering
\subfigure[]{\includegraphics[width = 0.42 \textwidth,page=3]{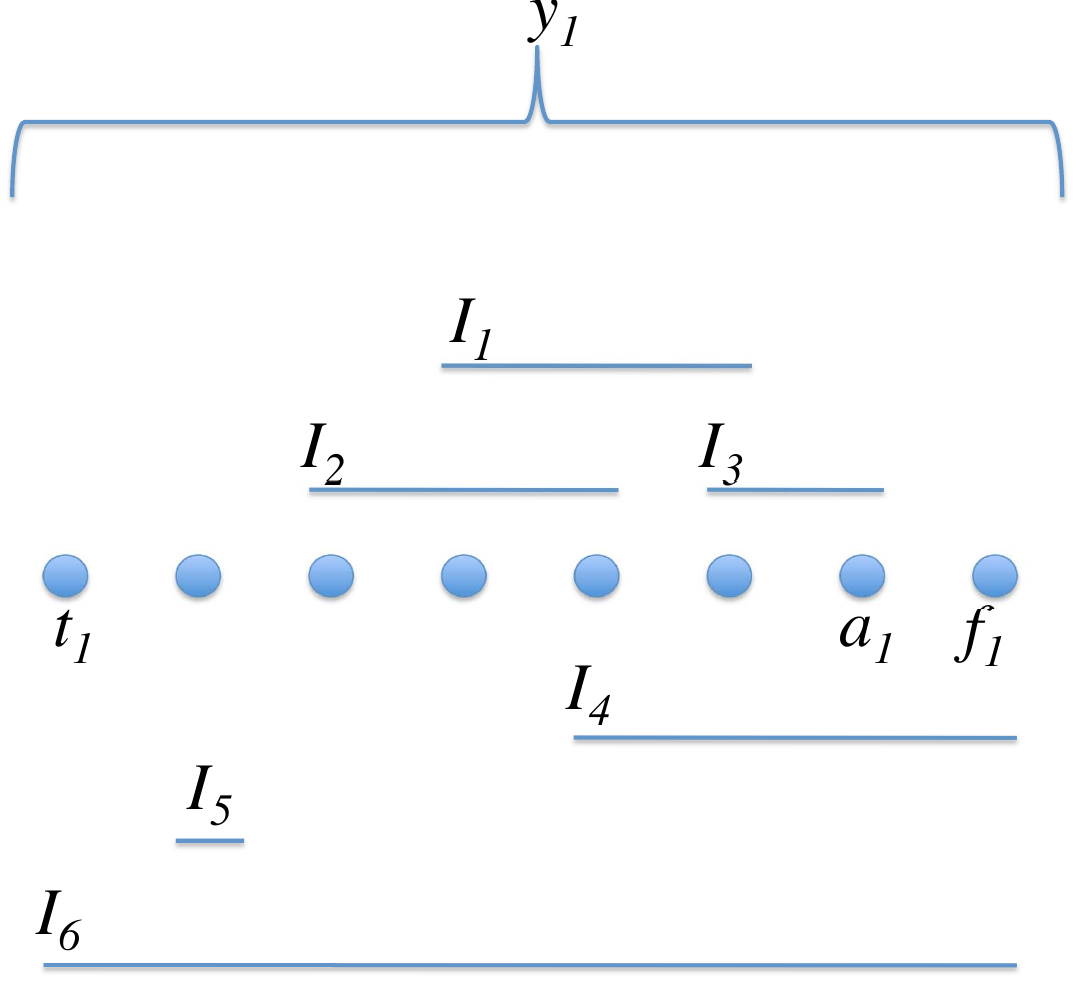}} \hspace*{0.05 \textwidth}
\subfigure[]{\includegraphics[width = 0.42 \textwidth,page=4]{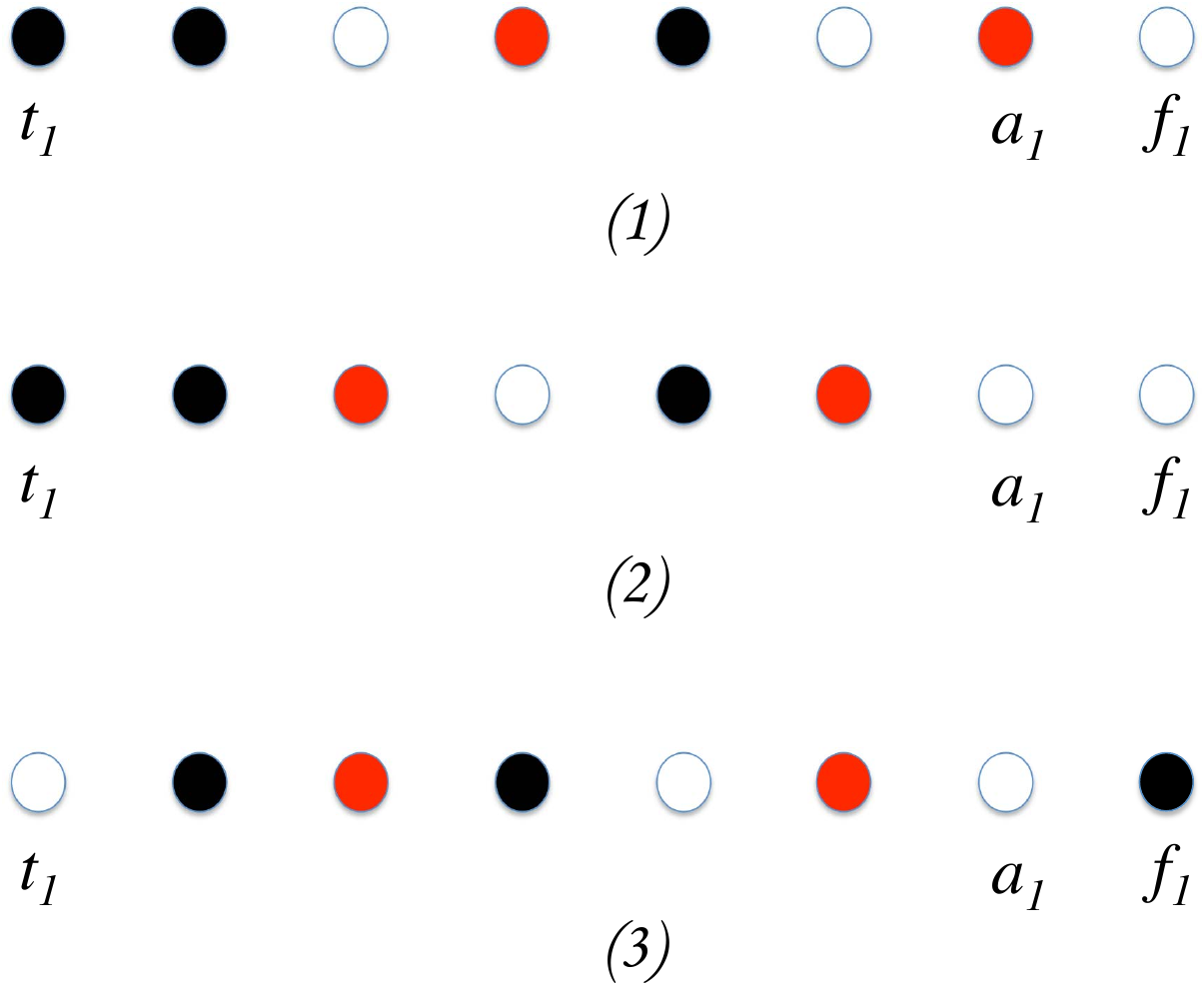}}
\caption{Literal $y_1$. The picture on the right shows the three feasible colorings. On a black and white 
printout red color appears as grey.}
\label{fig:2}
\end{figure}

\noindent The requirement function for such intervals changes whether $y_1=x_j$ or $y_1=\bar x_j$. If $y_1=x_j$, we let $r(I_1)= (1,1,1)$; $r(I_2)= (1,1,1)$; $r(I_3)= (1,0,1)$; $r(I_4)= (1,1,2)$; $r(I_5)= (0,1,0)$;  $r(I_6)= (2,3,3)$. For any feasible coloring there are only three possible outcomes for such sequence, reported in Fig.~\ref{fig:2}b.
Observe that the properties \eqref{(i)} and \eqref{(ii)} are enforced.

Now suppose that $y_1 = \bar x_j$:  then we switch the requirement function with respect to WHITE and BLACK, i.e. define it as follows: $r(I_1)= (1,1,1)$; $r(I_2)= (1,1,1)$; $r(I_3)= (1,1,0)$; $r(I_4)= (1,2,1)$; $r(I_5)= (0,0,1)$;  $r(I_6)= (2,3,3)$.
Trivially, the possible outcomes for such sequence are exactly the ones in Fig.~\ref{fig:2}b but exchanging the BLACK and WHITE colors.
 
\paragraph{Literal $y_3(i)$.}  The sequence of nodes representing literal $y_3$ is similar to the one representing $y_1$.  We still have a sequence of 8 nodes, and three special nodes $t_3,f_3$ and $a_3$.
As before, we let $t_3$ and $f_3$ encode the truth value of the variable $x_j$ that is referred to by $y_3$, while $a_3$ encodes the truth value of the literal $y_3$ itself. Therefore, we introduce a set $\mathcal I(y_3)$ of intervals in order to 
enforce the following properties:

\begin{property}\label{(iii)}
In any feasible coloring, exactly one among $t_3$ and $f_3$ will receive color BLACK. 
\end{property}

\begin{property}\label{(iv)}
In any feasible coloring, if $\chi(a_3) = RED$, then $\chi(t_3) = BLACK$ if $y_3 = x_j$, while $\chi(f_3) = BLACK$ if $y_3 = \bar x_j$. 
\end{property}

\noindent Fig.~\ref{fig:3}a shows the nodes and the six intervals that belong to $\mathcal I(y_3)$: observe that the sequence is similar to the one representing $y_1$,  but the position of node $a_3$  and the intervals are now ``mirrored''.
If $y_3 = \bar x_j$, we let $r(I_1)= (1,1,1)$;  $r(I_2)= (1,1,1)$;  $r(I_3)= (1,0,1)$;  $r(I_4)= (1,1,2)$;  $r(I_5)= (0,1,0)$;  $r(I_6)= (2,3,3)$.  
Fig.~\ref{fig:3}b reports the three possible outcomes for such sequence in a feasible coloring.  
Note that properties \eqref{(iii)} and \eqref{(iv)} hold.

Now suppose that $y_3 = x_j$: once again, we switch the requirement function with respect to WHITE and BLACK.

\begin{figure}[htb]
\centering
\subfigure[]{\includegraphics[width = 0.425 \textwidth,page=5]{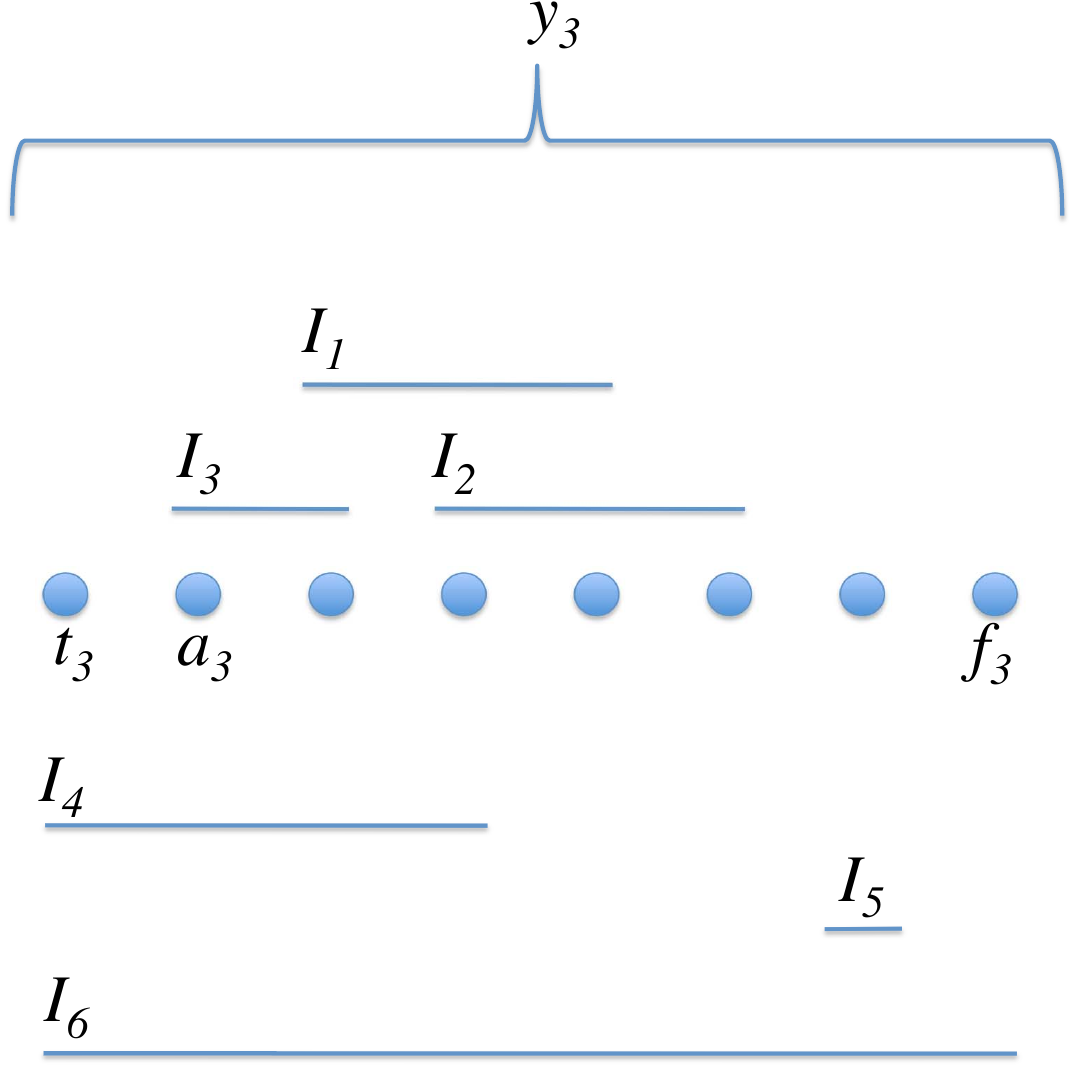}} \hspace*{0.05 \textwidth} 
\subfigure[]{\includegraphics[width = 0.425 \textwidth,page=6]{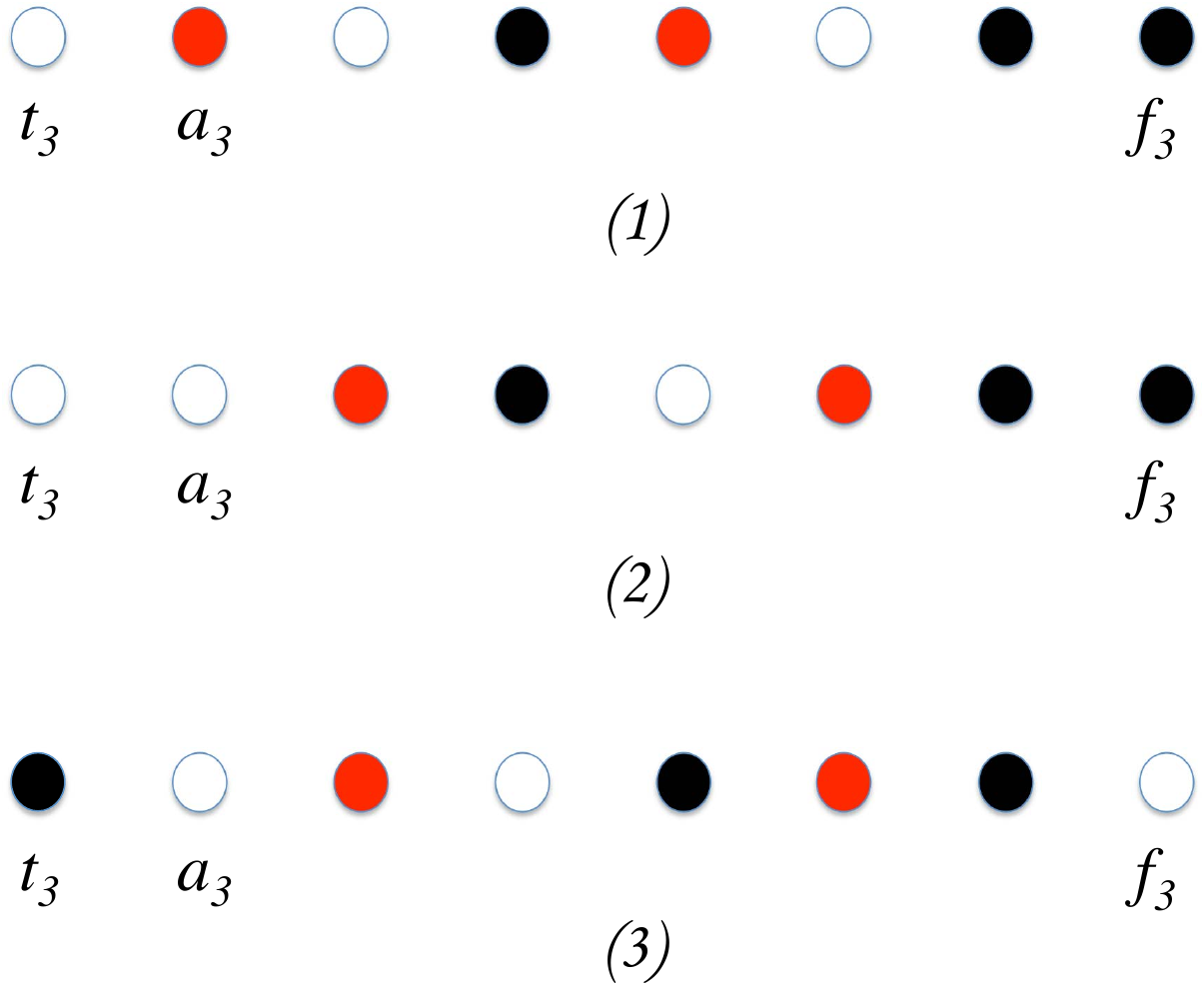}}
\caption{Literal $y_3$}
\label{fig:3}
\end{figure}

\paragraph{Literal $y_2(i)$.} The sequence of nodes representing literal $y_2$ is slightly more complicated. It is composed of 36 nodes, and among them there are 4 special nodes, namely $t_2,f_2,a_2^{\ell}$ and $a_2^r$ (see Fig.~\ref{fig:4}). 
Still, we let $t_2$ and $f_2$ encode the truth value of the variable $x_j$ that is referred to by $y_2$, while $a_2^{\ell}$ and $a_2^r$ encode the truth value of the literal. 

\begin{figure}[htb]
\centering
\includegraphics[width = 0.9 \textwidth,page=7]{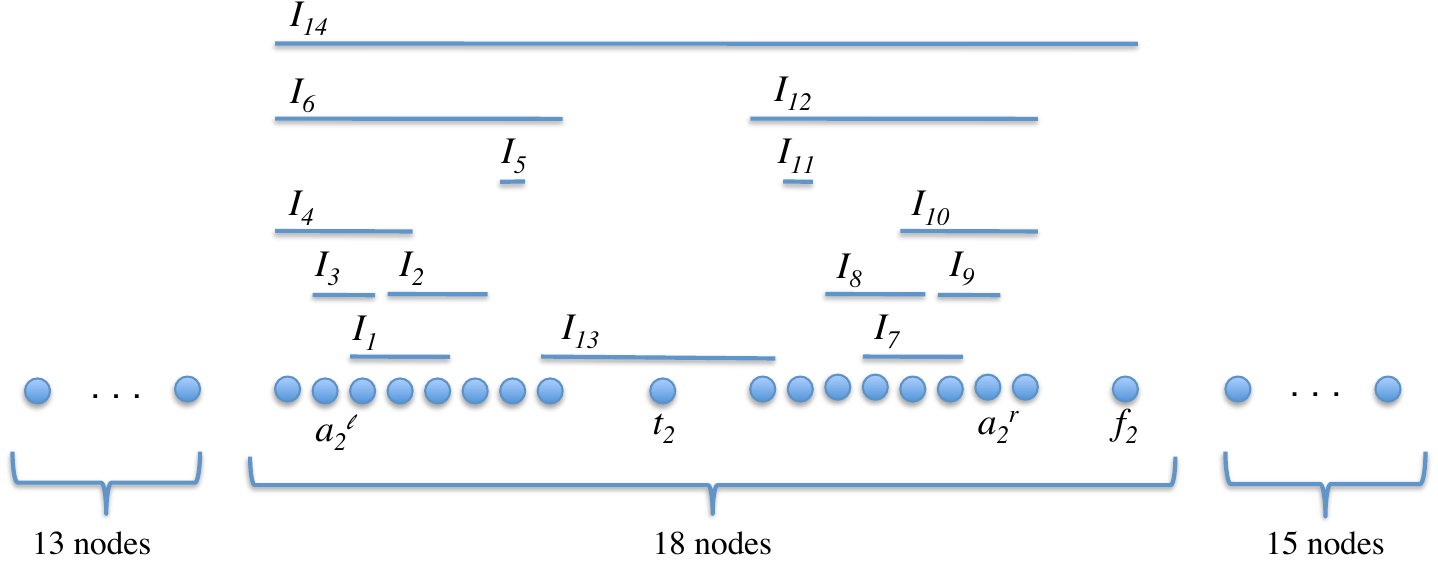} 
\caption{Literal $y_2$}
\label{fig:4}
\end{figure}

Similarly to the previous cases, we want to achieve the following goals: 1) given a feasible coloring, if $\chi(t_2)$ is BLACK, we want to be able to construct a truth assignment setting the variable $x_j$ to $true$, while if $\chi(f_2)$ is BLACK, we want to be able to construct a truth assignment setting the variable $x_j$ to $false$; 2) given a feasible coloring, if $\chi(a_2^{\ell})=\chi(a_2^r)=$ RED, we want to be able to construct a truth assignment where the literal $y_2$ is $true$. We are therefore interested in the following properties:

\begin{property}\label{(v)}
In any feasible coloring, exactly one among $t_2$ and $f_2$ will receive color BLACK.
\end{property}
 
\begin{property}\label{(vi)}
In any feasible coloring, if $\chi(a_2^{\ell}) = RED$ and $\chi(a_2^r) = RED$, then $\chi(t_2) = BLACK$ if $y_2 = x_j$, and $\chi(f_2) = BLACK$ if $y_2 = \bar x_j$. 
\end{property}

\noindent In this case, we introduce a set $\mathcal I(y_2)$ of 14 suitable intervals, shown in Fig.~\ref{fig:4}. 
The requirements for the case $y_2 = \bar x_j$ are given in the following table.
\[
\setlength{\arraycolsep}{1.75mm}
\begin{array}{c|cccccccccccccc}
          & I_1 & I_2 & I_3 & I_4 & I_5 & I_6 & I_7 &  I_8 & I_9 & I_{10} & I_{11} & I_{12} & I_{13} & I_{14} \\ \hline
RED     & 1 & 1 & 1 & 1 & 0 & 2 & 1 & 1 & 1 & 1 & 0 & 2 & 0 & 4 \\
BLACK & 1 & 1 & 0 & 1 & 1 & 3 & 1 & 1 & 0 & 1 & 1 & 3 & 2 & 7 \\
WHITE & 1 & 1 & 1 & 2 & 0 & 3 & 1 & 1 & 1 & 2 & 0 & 3 & 1 & 7 
\end{array}
\]

\noindent Observe that the set of intervals $\{I_1, \dots, I_6 \}$ is defined exactly as the set $\mathcal I(y_3)$, therefore the possible outcomes for the sequence of 8 nodes covered by such intervals are as in Fig.~\ref{fig:3}b.  
Similarly, the set of intervals $\{I_7, \dots, I_{12} \}$ is defined exactly as the set $\mathcal I(y_1)$, therefore the possible outcomes for the sequence of 8 nodes covered by such intervals are as in Fig.~\ref{fig:2}b.  
Combining $r(I_6)$ and $r(I_{12})$ with $r(I_{14})$, it follows that in any feasible coloring $\chi$, exactly one node among $t_2$ and $f_2$ has WHITE (resp. BLACK) color, 
enforcing Property $\eqref{(v)}$.
Still, note that if $\chi(a_2^{\ell}) = RED$ and $\chi(a_2^r) = RED$, then both the leftmost node and the rightmost node covered by interval $I_{13}$ have color BLACK, therefore $t_2$ must have color WHITE otherwise $r(I_{13})$ is violated. Together with Property  \eqref{(v)}, this enforces Property \eqref{(vi)}.  

In case $y_2 = x_j$, once again we switch the requirement function with respect to WHITE and BLACK. 

\smallskip
\noindent It remains to describe the role played by the first 13 nodes and the last 15 nodes of the sequence, that so far we did not consider. We are going to do it in the next paragraph.

\paragraph{Intervals encoding truth values of literals.} For each clause $C_i$, we add another set $\mathcal I(C_i)$ of intervals, in order to link the nodes encoding the truth values of its three literals. The main goal we pursue is the following: given a feasible coloring, we want to be able to construct a truth assignment such that at least one of the three literals is $true$. To achieve this, already having properties \eqref{(ii)}, \eqref{(iv)} and \eqref{(vi)}, we only need the following property:

\begin{property}\label{(vii)}
For any feasible coloring, if $\chi(a_1) \neq RED$ and $\chi(a_3) \neq RED$ , then $\chi(a_2^{\ell}) = \chi(a_2^r) = RED$.
\end{property}

\noindent Fig.~\ref{fig:5} shows the six intervals that belong to $\mathcal I(C_i)$. The requirement function is: $r(I_1)=(1,2,2);$ $r(I_2)=(1,2,2);$ $r(I_3)=(1,6,6);$ $r(I_4)=(1,3,3);$ $r(I_5)=(1,2,2);$ $r(I_6)=(1,7,7)$. We now show that Property \eqref{(vii)} holds.
Suppose $\chi$ is a feasible coloring, and let $v_1, \dots, v_{13}$ be the first 13 nodes of the sequence introduced for literal $y_2$. By construction, if $\chi(a_1) \neq RED$, then there is a node $v_j : \chi(v_j) = RED$ and $j \in \{1,2,3\}$ , otherwise $r(I_1)$ is violated. Similarly, if $\chi(a_2^{\ell}) \neq RED$, then there is a node $v_j : \chi(v_j) = RED$ and $j \in \{11,12,13\}$ , otherwise $r(I_2)$ is violated. On the other hand, this subsequence contains exactly one node with RED color, otherwise $r(I_3)$ is violated. It follows that at least one among $a_1$ and $a_2^{\ell}$ has RED color. The same conclusions can be stated for nodes $a_2^r$ and $a_3$. Putting all together, it follows that the Property \eqref{(vii)} holds. 

\begin{figure}[htb]
\centering
\includegraphics[width = 0.9 \textwidth,page=8]{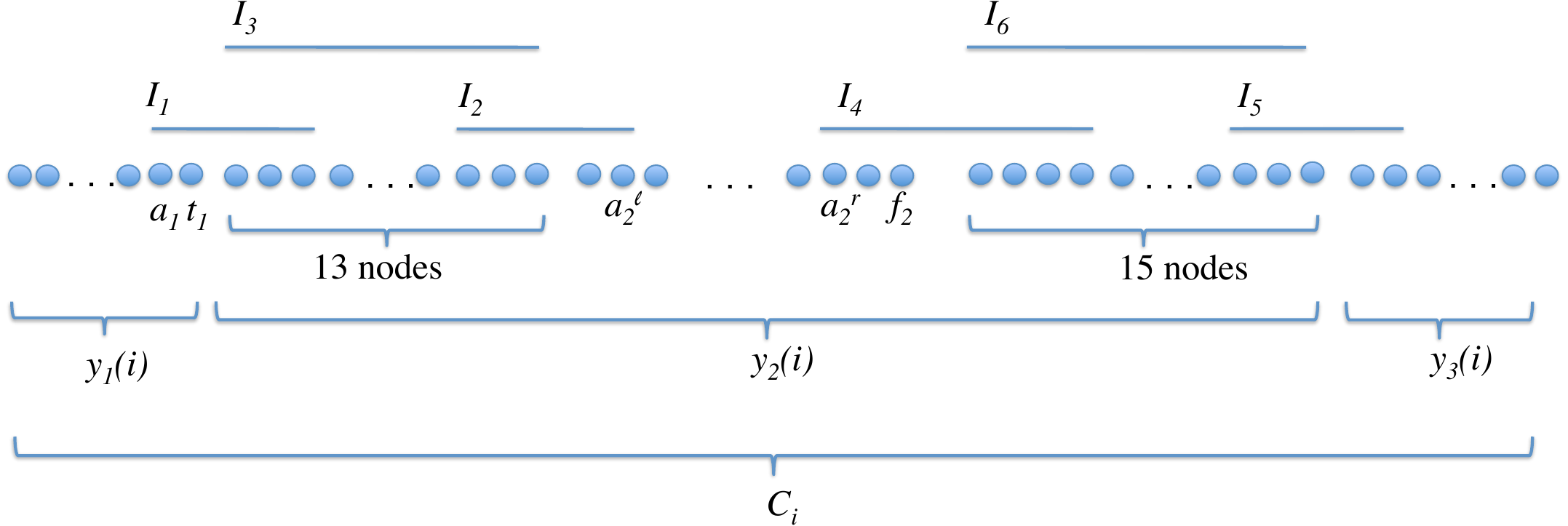} 
\caption{Set of intervals $\mathcal I(C_i)$.}
\label{fig:5}
\end{figure}

\paragraph{Intervals encoding truth value of variables (later also called: variable intervals).} Our last set of intervals will force different nodes to take the same color, if they encode the truth value of the same variable. In particular, we aim at having the following property:

\begin{property}\label{(viii)}
In any feasible coloring, $\chi(t_h(i) ) = \chi(t_k(i'))$ if both literals $y_h(i)$ and $y_k(i')$ refer to the same variable $x_j$.
\end{property}

\noindent To achieve this, for each pair of such literals we add a big interval $I(y_h(i), y_k(i'))$ from $f_k(i')$ to $t_h(i)$ (assuming $i' < i$ without loss of generality). Note that, by construction, there is a subset of intervals that partitions all the internal nodes covered by the interval. That means, we know exactly the number of such nodes that must be colored with color RED, BLACK and WHITE (say $z_1,z_2,z_3$ respectively). Then, we let the requirement function be $r(I(y_h(i), y_k(i'))) = (z_1, z_2+1, z_3+1)$. Under these assumptions, if $\chi$ is a feasible coloring then  $\chi(t_h(i)) \neq \chi(f_k(i'))$, and in particular one node will have WHITE color and the other one BLACK color. Combining this with properties \eqref{(i)},\eqref{(iii)} and \eqref{(v)}, the result follows.

\medskip
Notice that such an interval constrained 3-coloring instance can clearly be constructed in polynomial time. Now we discuss the following claim in more details.

\begin{claim}
There exists a truth assignment satisfying the 3-SAT instance if and only if there exists a feasible coloring $\chi$ for the interval constrained 3-coloring instance.
\end{claim}

First, suppose there exists a feasible coloring. We construct a truth assignment as follows. We set a variable $x_j$ to $true$ if $\chi(t_h(i))=BLACK$, and to $false$ otherwise, where $y_h(i)$ is any literal referring to $x_j$. 
Note that, by Property  \eqref{(viii)}, the resulting truth value does not depend on the literal we take. 
Still, combining Property \eqref{(vii)} with properties \eqref{(ii)},\eqref{(iv)} and \eqref{(vi)}, we conclude that, for each clause, at least one literal will be $true$. By construction, we therefore end up with a truth assignment satisfying the 3-SAT instance. The result follows.

\smallskip
Now suppose that there is a truth assignment satisfying the 3-SAT instance. The basic idea, is to construct a coloring $\chi$ such that the following property holds for all literals:

\begin{property}\label{(viv)}
$\chi(t_h(i)) =$ BLACK (resp. WHITE) if and only if $y_h(i)$ refers to a $true$-variable (resp. $false$-variable).
\end{property}

\smallskip
\noindent Consider the sequence of nodes representing literal $y_1(i)$, and suppose $y_1(i) = x_j$ for some $j$. We color such nodes as in Fig.~\ref{fig:2}b-\emph{(1)} if the literal is $true$ in the truth assignment, and as in Fig.~\ref{fig:2}b\emph{-(3)} otherwise. If $y_1(i) = \bar {x}_j$, switch BLACK and WHITE colors, in both previous cases. Now focus on the sequence of nodes representing literal $y_3(i)$. If $y_3(i) =\bar x_j$ for some $j$, we color such nodes as in Fig.~\ref{fig:3}b-\emph{(1)} if the literal is $true$, and as in Fig.~\ref{fig:3}b\emph{-(3)} otherwise. If $y_3(i) = {x}_j$, switch BLACK and WHITE colors, in both previous cases. Finally, consider the sequence of nodes representing literal $y_2(i)$. Suppose $y_2(i) = \bar x_j$. We color the 18 nodes in the middle of the sequence as in Fig.~\ref{fig:8}\emph{-(1)} if $y_2(i)$ is $true$, as in Fig.~\ref{fig:8}\emph{-(2)} if both $y_2(i)$ and $y_1(i) $ are $false$, and as in Fig.~\ref{fig:8}\emph{-(3)}  otherwise. Once again, if $y_2(i) = {x}_j$, we switch BLACK and WHITE colors, in all the previous three cases. Notice that, 
by construction, Property \eqref{(viv)} holds, and all requirements for the intervals in $\mathcal I(y_h(i))$ $(i=1, \dots, q$ and $h=1, 2, 3$) are not violated.

\begin{figure}[htb]
\centering
\includegraphics[width = 0.9 \textwidth,page=9]{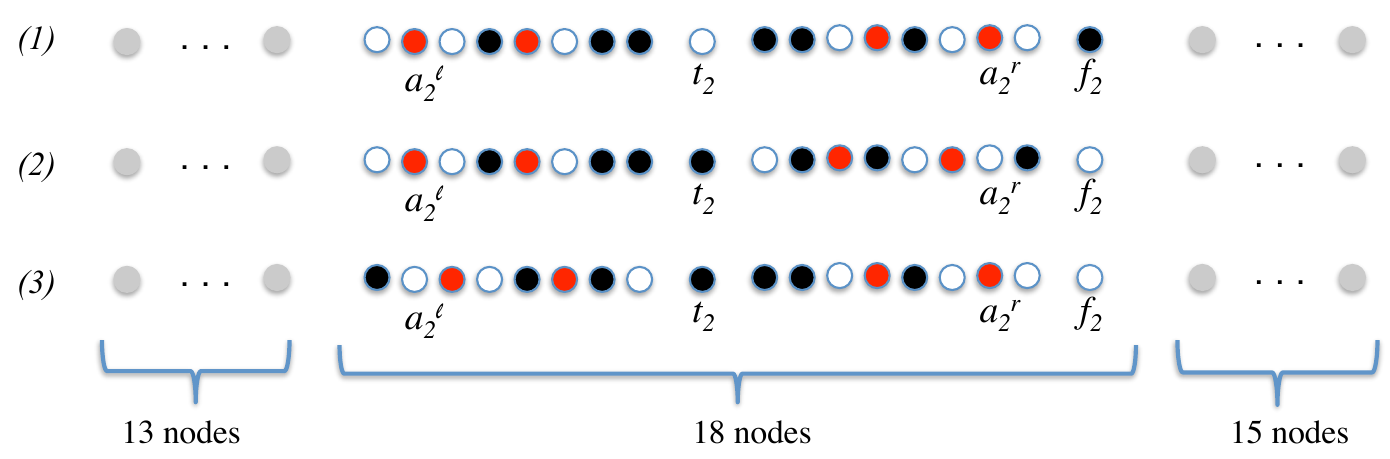} 
\caption{Coloring of nodes representing literal $y_2$}
\label{fig:8}
\end{figure}

Now we show how to color the first 13 nodes $(v_1, \dots, v_{13})$ and the last 15 nodes $(w_1, \dots, w_{15})$ of the sequence representing literal $y_2(i)$, in such a way that the requirements of the intervals $I_1, \dots, I_6$ in $\mathcal I(C_i)$ are not violated ($i=1, \dots, q)$. Note that, by construction, at least one node among $a_1$ and $a_2^{\ell}$ is colored with RED. In fact, if $y_1(i)$ is $true$ then $\chi(a_1) =$ RED, while if $y_1(i) = false$ then $a_2^{\ell}$ is colored with RED. Similarly, at least one node among $a_3$ and $a_2^r$ is colored with RED, since $\chi(a_2^r) \neq RED$ only if both literals $y_1(i)$ and $y_2(i)$ are $false$: then, necessarily $y_3(i)$ is $true$, and therefore $\chi(a_3) = RED$. 
Let us focus on the nodes $v_1, \dots, v_{13}$, and let $u$ be the node in between $v_{13}$ and $a_2^{\ell}$. In the following, we refer to WHITE as the \emph{opposite} color of BLACK and vice versa. 
As we already discuss, we can have only two cases:

Case 1: $\chi(a_1) = \chi(a_2^{\ell}) = RED$. We color $v_1$ with the opposite color of $f_1$, and the nodes $v_2$ and $v_3$ with BLACK and WHITE. Note that $r(I_1)$ is not violated. We then color $v_4,v_5,v_6$ with the opposite color of $v_1,v_2,v_3$ respectively. 
Similarly, we color $v_{13}$ with the opposite color of $u$. Then, we color $v_{12}$ and $v_{11}$ with BLACK and WHITE, so that $r(I_2)$ is not violated.  
Once again, we assign to $v_{10},v_9,v_8$ the opposite color of $v_{13},v_{12},v_{11}$ respectively. Finally, we let $\chi(v_7) = RED$. Note that $r(I_3)$ is not violated.

Case 2: $\chi(a_1) \neq RED$ and $\chi(a_2^{\ell}) = RED$, or vice versa. Suppose $\chi(a_1) \neq RED$ (the other case is similar). Both nodes $a_1$ and $f_1$ can have only BLACK or WHITE colors. Then, we can color $v_1$ and $v_2$ with the opposite color of $a_1$ and $f_1$ respectively, and $v_3$ with color RED, so that $r(I_1)$ is not violated. Still, we color
$v_4$ and $v_5$ with the opposite color of $v_1$ and $v_2$. Finally, we color $v_6$ and $v_7$ with BLACK and WHITE. To the remaining nodes $v_8, \dots, v_{13}$ we assign the same colors as in Case 1. One checks that requirements of intervals $I_2$ and $I_3$ are not violated. 

One can prove in a similar manner that nodes $(w_1, \dots, w_{15})$ can be properly colored, without violating the requirements of intervals $I_4,I_5,I_6$.

\smallskip
Finally, since Property \eqref{(viv)} holds, it is easy to see that, for each couple of literals $y_h(i), y_k(i')$, the requirement $r(I(y_h(i), y_k(i')))$ is also not violated. The result then follows.%\qed

\end{proof}

\section{Gap hardness}

We will now argue that not only the interval constrained 3-coloring problem
but also its gap version is NP-hard, i.e., it is hard to distinguish between
satisfiable instances and those where only up to a $(1-\epsilon)$ fraction
of constraints may be simultaneously satisfied.

For the purpose of our argument we will use the following, rather restricted, 
definition of gap hardness. We will only talk about 
maximization versions of constraint satisfaction problems.
Think of an instance of the problem as being equipped with an
additional parameter $t$ called threshold. We ask for a polynomial
time algorithm which given the instance answers: 
\begin{itemize}
\item
``YES'' if all the constraints
can be satisfied, 
\item
``NO'' if there is no solution satisfying
more than $t$ constraints.
\end{itemize}
Note that for instances, where more than $t$ but not all constraints
can be simultaneously satisfied, any answer is acceptable.
We will now restrict our attention to the case where the threshold is a fixed fraction
of the total amount of constraints in the instance. 
We call problem A to be \emph{gap NP-hard} if there exists a positive $\epsilon$
such that there is no polynomial time algorithm to separate feasible instances
from those where only at most a $(1-\epsilon)$ fraction of the constraint can be simultaneously
satisfied unless $P = NP$.

Observe that gap NP-hardness implies APX-hardness, but not vice versa.
For example the linear ordering problem (also known as max-subdag) is APX-hard~\cite{papa_apx}, 
but is not gap NP-hard, since feasible instances may be found by topological sorting.

Let us first note that the 3-SAT problem, which we used in the reduction
from the previous section, has the gap hardness property. It is the essence
of the famous PCP theorems that problems with such gap hardness exist.
For a proof of the gap hardness of 3-SAT see~\cite{gap_amp}.

Before we show how to modify our reduction to prove gap hardness of the interval constraint coloring problem,
we need to introduce the notion of \emph{expander graphs}. 
For brevity we will only give the following extract from~\cite{gap_amp}.

\begin{definition}
  Let $G = (V,E)$ be a $d$-regular graph. Let $E(S,\overline{S}) = | (S\times\overline{S}) \cap E |$
  equal the number of edges from a subset $S \subseteq V$ to its complement. The \emph{edge expansion}
  of $G$ is defined as
  \[
  h(G)= \min_{S:|S|\leq |V|/2}\frac{E(S,\overline{S})}{|S|}.
  \]
\end{definition}

\begin{lemma}
  There exists $d_0 \in \setN$ and $h_0 > 0$, such that there is a polynomial-time constructible family
  $\{ X_n \} _{n \in \setN}$ of $d_0$-regular graphs $X_n$ on $n$ vertices with $h(X_n) \geq h_0$. (Such graphs
  are called expanders).
\end{lemma}

Let us now give a ``gap preserving'' reduction from gap 3-SAT to 
gap interval constrained 3-coloring. Consider the reduction from the previous section.
Observe that the amount of intervals in each literal gadget, and therefore also in each clause gadget, is constant.
The remaining intervals are the variable intervals. While it is sufficient for 
the NP-hardness proof to connect occurrences of the same variable in a ``clique'' fashion with
variable intervals, it produces a potentially quadratic number of intervals.
Alternatively, one could connect these occurrences in a ``path'' fashion,
but it would give too little connectivity for the gap reduction.
The path-like connection has the desired property of using only linear amount of intervals,
since each occurrence of a variable is linked with at most two other ones.
We aim at providing more connectivity while not increasing the amount of intervals too much.
A perfect tool to achieve this goal is a family of expander graphs.

Consider the instance of the interval coloring problem obtained by the reduction
from the previous section, but without any variable intervals yet.
Consider literal gadgets corresponding to occurrences of a particular variable $x$.
Think of these occurrences as of vertices of a graph $G$. Take an expander graph $X_{|V(G)|}$ 
and connect two occurrences of $x$ if the corresponding vertices in the expander are connected. 
For each such connection use a pair of intervals. These intervals should be the
original variable interval and an interval that is one element shorter on each of the sides.
We will call this pair of intervals a variable link.
Repeat this procedure for each of the variables.

Observe that the number of variable links that we added is linear
since all the used expander graphs are $d_0$-regular.
By contrast to the simple path-like connection, we now have the property,
that different occurrences of the same variable have high edge connectivity.
This can be turned into high penalty for inconsistent valuations 
of literals in an imperfect solution.  

\begin{theorem}
Constrained interval 3-coloring is gap NP-hard.
\end{theorem}

\begin{proof}
We will argue that the above described reduction is a gap-preserving reduction from
the gap 3-SAT problem to the gap interval 3-coloring problem. We need to prove that
there exists a positive $\epsilon$ such that feasible instances are hard to separate from
those less than $(1-\epsilon)$ satisfiable.

Let $\epsilon_0$ be the constant in the gap hardness of gap 3-SAT.
We need to show two properties: that the ``yes'' instances of the gap 3-SAT problem
are mapped to ``YES'' instances of our problem, and also that the ``NO''
instances are mapped to ``NO'' instances. 

The first property is simple, already in the NP-hardness proof in the previous section
it was shown that feasible instances are mapped by our reduction into feasible ones.  
To show the second property, we will take the reverse direction and argue that
an almost feasible solution to the coloring instance can be transformed into an almost feasible
solution to the SAT instance.

Suppose we are given a coloring $\chi$ that violates at most $\epsilon$ fraction of the constraints.
Suppose the original 3-SAT instance has $q$ clauses, then our interval coloring instance has at most $c \cdot q$
intervals for some constant $c$. The number of unsatisfied intervals in the coloring $\chi$ is then at most $\epsilon q c$.

We will say that a clause is \emph{broken} if at least one of the intervals encoding it is not satisfied by $\chi$. 
We will say that a variable link is broken if one of its intervals is not satisfied or one of the clauses it connects is broken. An unsatisfied variable link interval contributes a single broken link; an unsatisfied interval within a clause breaks at most $3 d_0$ intervals connected to the clause. Therefore, there is at most $3 d_0 \epsilon q c$ broken
variable links in total.

Recall that each variable link that is not broken connects
occurrences of the same variable in two different not broken clauses. 
Moreover, by the construction of the variable link, these two occurrences 
display the same logical value of the variable.

Consider the truth assignment $\phi$ obtained as follows. For each variable consider its occurrences
in the not broken clauses. Each occurrence associates a logical value to the variable.
Take for this variable the value that is displayed in the bigger set of not broken clauses,
break ties arbitrarily.

We will now argue, that $\phi$ satisfies a big fraction of clauses.
Call a clause \emph{bad} if it is not broken, but it contains a literal
such that in the coloring $\chi$ this literal was active, but $\phi$ evaluates
this literal to false. Observe that if a clause is neither broken nor bad, then
it is satisfied by $\phi$. It remains to bound the amount of bad clauses.

Consider the clauses that become bad from the choice of a value that $\phi$
assigns to a particular variable $x$. Let $b_x$ be the number of such clauses. 
By the connectivity property of expanders,
the amount of variable links connecting these occurrences of $x$ with other occurrences
is at least $h_0 b_x$. As we observed above, all these variable links are broken.
Since there are in total at most $3 d_0 \epsilon q c$ broken links, we obtain that 
there is at most $\tfrac{3}{h_0} d_0 \epsilon q c$ bad clauses. Hence, there are at most
$(\tfrac{3}{h_0} d_0 +1)\epsilon q c$ clauses that are either bad or broken and they cover
all the clauses not satisfied by $\phi$.

It remains to fix $\epsilon = \tfrac{h_0}{(3 d_0 + h_0)c} \epsilon_0$ to obtain the property,
that more than $\epsilon_0$ unsatisfiable instances of 3-SAT are mapped to more than
$\epsilon$ unsatisfiable instances of the constrained interval 3-coloring problem.
%\qed
\end{proof}

\smallskip
\noindent
{\bf Acknowledgment}

\noindent
We thank Steven Kelk for valuable discussion. 

\bibliographystyle{splncs}
\bibliography{interval3coloring}

\end{document}